\documentclass[prb,twocolumn,showpacs]{revtex4-1}

\usepackage{graphicx}
\usepackage{amsmath}
\usepackage{amssymb}
\usepackage{epsfig}
\usepackage{epstopdf}
\usepackage{gensymb}
\usepackage[sort&compress]{natbib}

\def\lsim{\lower -0.3ex \hbox{$<$} \kern -0.75em \lower 0.7ex \hbox{$\sim$}}
\def\gsim{\lower -0.3ex \hbox{$>$} \kern -0.75em \lower 0.7ex \hbox{$\sim$}}

\def\vare{\varepsilon}

\begin{document}

\title{Hofstadter's butterfly in moir\'{e} superlattices: A fractal quantum Hall effect}

\author{C. R. Dean$^{1,2}$}
\author{L. Wang$^{2}$}
\author{P. Maher$^{3}$}
\author{C. Forsythe$^{3}$}
\author{F. Ghahari$^{3}$}
\author{Y. Gao$^{2}$}
\author{J. Katoch$^{4}$}
\author{M. Ishigami$^{4}$}
\author{P. Moon$^{5}$}
\author{M. Koshino$^{5}$}
\author{T. Taniguchi$^{6}$}
\author{K. Watanabe$^{6}$}
\author{K. L. Shepard$^{1}$}
\author{J. Hone$^{2}$}
\author{P. Kim$^{3}$}

\affiliation{$^{1}$Department of Electrical Engineering, Columbia University, New York, NY}
\affiliation{$^{2}$Department of Mechanical Engineering, Columbia University, New York, NY}
\affiliation{$^{3}$Department of Physics, Columbia University, New York, NY}
\affiliation{$^{4}$Department of Physics and Nanoscience Technology Center, University of Central Florida, Orlando, FL}
\affiliation{$^{5}$ Department of Physics, Tohoku University, Sendai, Japan}
\affiliation{$^{6}$National Institute for Materials Science, 1-1 Namiki, Tsukuba, Japan}

\pacs{xx}
\maketitle

\textbf{Electrons moving through a spatially periodic lattice potential develop a quantized energy spectrum consisting of discrete Bloch bands. In two dimensions, electrons moving through a magnetic field  also develop a quantized energy spectrum, consisting of  highly degenerate Landau energy levels. In 1976 Douglas Hofstadter theoretically considered the intersection of these two problems and discovered that  2D electrons subjected to both a magnetic field and a periodic electrostatic potential exhibit a self-similar recursive energy spectrum\cite{Hofstadter:1976}.  Known as Hofstadter's butterfly, this complex spectrum results from a delicate interplay between the characteristic lengths associated with the two quantizing fields\cite{Hofstadter:1976,Wannier:1978,Claro:1979,Streda:1982b,Thouless:1982,MacDonald:1983,Koshino:2006,Nemec:2007,Bistritzer:2011,Moon:2012, Beugeling:2012}, and represents one of the first quantum fractals discovered in physics. In the decades since, experimental attempts to study this effect have been limited by difficulties in reconciling the two length scales.
Typical crystalline systems ($<1$~nm periodicity) require impossibly large magnetic fields to reach the commensurability condition, while in artificially engineered structures ($\gtrsim100$~nm), the corresponding fields are too small to completely overcome disorder\cite{Weiss:1990,Gerhardts:1991,Albrecht:1998,Nakamura:1998,Albrecht:1999,Schlosser:1999,Albrecht:2003,Geisler:2004, Melinte:2004}. Here we demonstrate that moir\'{e} superlattices arising in bilayer graphene coupled to hexagonal boron nitride provide a nearly ideal-sized periodic modulation, enabling unprecedented experimental access to the fractal spectrum.  We confirm that quantum Hall effect features associated with the fractal gaps are described by two integer topological quantum numbers, and report evidence of their recursive structure.  Observation of Hofstadter's spectrum in graphene provides the further opportunity to investigate emergent behaviour within a fractal energy landscape in a system with tunable internal degrees of freedom.}

The total number of electron states per area of a completely filled Bloch band is exactly $n_0=1/A$, where $A$ is the area of the unit cell of the periodic potential. In a magnetic field, the number of states per area of each filled Landau level (LL) is given by $B/\phi_0$ where $\phi_0=h/e$ is the magnetic flux quanta. The quantum description of electrons subjected simultaneously to both a periodic electric field and a magnetic field can be simply parameterized by the dimensionless ratio $\phi/\phi_0$ where $\phi=BA$ is magnetic flux per unit cell.  The general solution of this problem, however, exhibits a rich complexity due to the incommensurate periodicities between the Bloch and Landau states\cite{Harper:1955}. In his seminal work\cite{Hofstadter:1976}, Hofstadter showed that for commensurate fields, corresponding to rational values of $\phi/\phi_{0}=p/q$, where  $p$ and $q$ are co-prime integers, the single-particle Bloch band splits into $q$ subbands (beginning with the Landau level description it is  equivalently shown\cite{Wannier:1978} that  at these same rational values each Landau level splits into p subbands). This results in a quasi-continuous distribution of incommensurate quantum states that exhibits self-similar recursive structure, yielding the butterfly-like fractal energy diagram (see SI).

Important insight into this problem was subsequently provided by Wannier\cite{Wannier:1978}, who considered  the density of charge carriers, $n$, required to fill each fractal subband. Replotting the Hofstadter energy spectrum as integrated density versus field, Wannier realized that all spectral gaps are constrained to linear trajectories in the density-field diagram. This can be described by a simple Diophantine relation

\begin{equation}
%\label{eqn:Diophantine}
(n/n_{o})=t({\phi}/{\phi_{o}})+s
\end{equation}

\noindent
where $n/n_{o}$ and $\phi/\phi_{o}$ are the normalized carrier density and magnetic flux, respectively, and $s$ and $t$ are both integer valued. Here $n/n_{o}$ represents the Bloch band filling fraction, which is distinct from the usual Landau level filling fraction, $\nu=n\phi_0/B$ (the two are related by the normalized flux, i.e. $n/n_{o}=\nu\phi/\phi_{o})$. The physical significance of the $s$ and $t$ quantum numbers became fully apparent with the discovery of the integer quantum Hall effect\cite{Klitzing:1980} in 1980, after which it was shown, simultaneously by both St\v{r}reda\cite{Streda:1982b} and Thouless \text{et al}\cite{Thouless:1982}, that the Hall conductivity associated with each mini-gap in the fractal spectrum is quantized according to $\sigma_{xy}=t e^{2}/h$.  The second quantum number, $s$, physically corresponds to the Bloch band filling index in the fractal spectrum\cite{MacDonald:1983}. This formalism suggests several unique and unambiguous experimental signatures associated with the Hofstadter energy spectrum that are distinct from the conventional quantum Hall effect: i) the Hall conductance can vary non-monotonically and can even fluctuate in sign, ii) Hall conductance plateaus together with vanishing longitudinal resistance can appear at \textit{non-integer} LL filling fractions, iii) the Hall conductance plateau remains quantized to integral multiples of $e^{2}/h$, however, the quantization integer is not directly associated with the usual LL filling fraction. Instead, quantization is equal to  the slope of the gap trajectory in the $n/n_o$ versus $\phi/\phi_o$ Wannier diagram, in accordance with the Diophantine equation.

Mini-gaps within the fractal energy spectrum become significant only once the magnetic length ($l_{B}=\sqrt{\hbar/eB}$), which characterizes the cyclotron motion, is of the same order as the wavelength of the periodic potential, which characterizes the Bloch waves.  For usual crystal lattices, where the inter-atomic spacing is a few \r{a}ngstroms, the necessary magnetic field is impossibly large, in excess of 10,000~T.  The main experimental effort therefore has been to lithographically define artificial superlattices\cite{Weiss:1990,Gerhardts:1991,Albrecht:1998,Nakamura:1998,Albrecht:1999,Schlosser:1999,Albrecht:2003,Geisler:2004, Melinte:2004} with unit cell dimension of order tens of nanometers so that the critical magnetic field remains small enough to be achievable in the lab, yet still large enough so that the quantum Hall effect is fully resolved without being smeared out by disorder.  Fabricating the optimally-sized periodic lattice, while maintaining coherent registry over the full device and without introducing substantial disorder has proven a formidable technical challenge. Patterned GaAs/AlGaAs heterostructures with $\sim100$~nm periodic gates provided the first experimental support for Hofstadter's predictions\cite{Schlosser:1999,Albrecht:2003,Geisler:2004}.  However, limited ability to tune the carrier density or reach the fully developed quantum Hall effect (QHE) regime in these samples has made it difficult to map out the complete spectrum.  While similar concepts have also been pursued in non-solid-state model systems\cite{Kuhl:1998,Jaksch:2003}, the  rich physics of the Hofstadter spectrum remains largely unexplored.

Heterostructures consisting of atomically thin materials in a multi-layer stack provide a new route towards realizing a two-dimensional system with laterally modulated periodic structure.  In particular, coupling between graphene and hexagonal boron nitride (hBN), whose crystal lattices are isomorphic, results in a periodic moir\'{e} pattern.  The moir\'{e} wavelength is directly related to the angular rotation between the two lattices\cite{Xue:2011,Decker:2011,Yankowitz:2012}, and is tunable through the desired length scales without the need for lithographic techniques\cite{Bistritzer:2011,Moon:2012}.  Moreover hBN provides an ideal substrate for achieving high mobility graphene devices, crucial for high resolution quantum Hall measurements\cite{Dean:2010,Dean:2011}, while field effect gating in graphene allows the Fermi energy to be continuously varied through the entire moir\'{e} Bloch band. 

\begin{figure}[t]
	\begin{center}
	\includegraphics[width=1\linewidth,angle=0,clip]{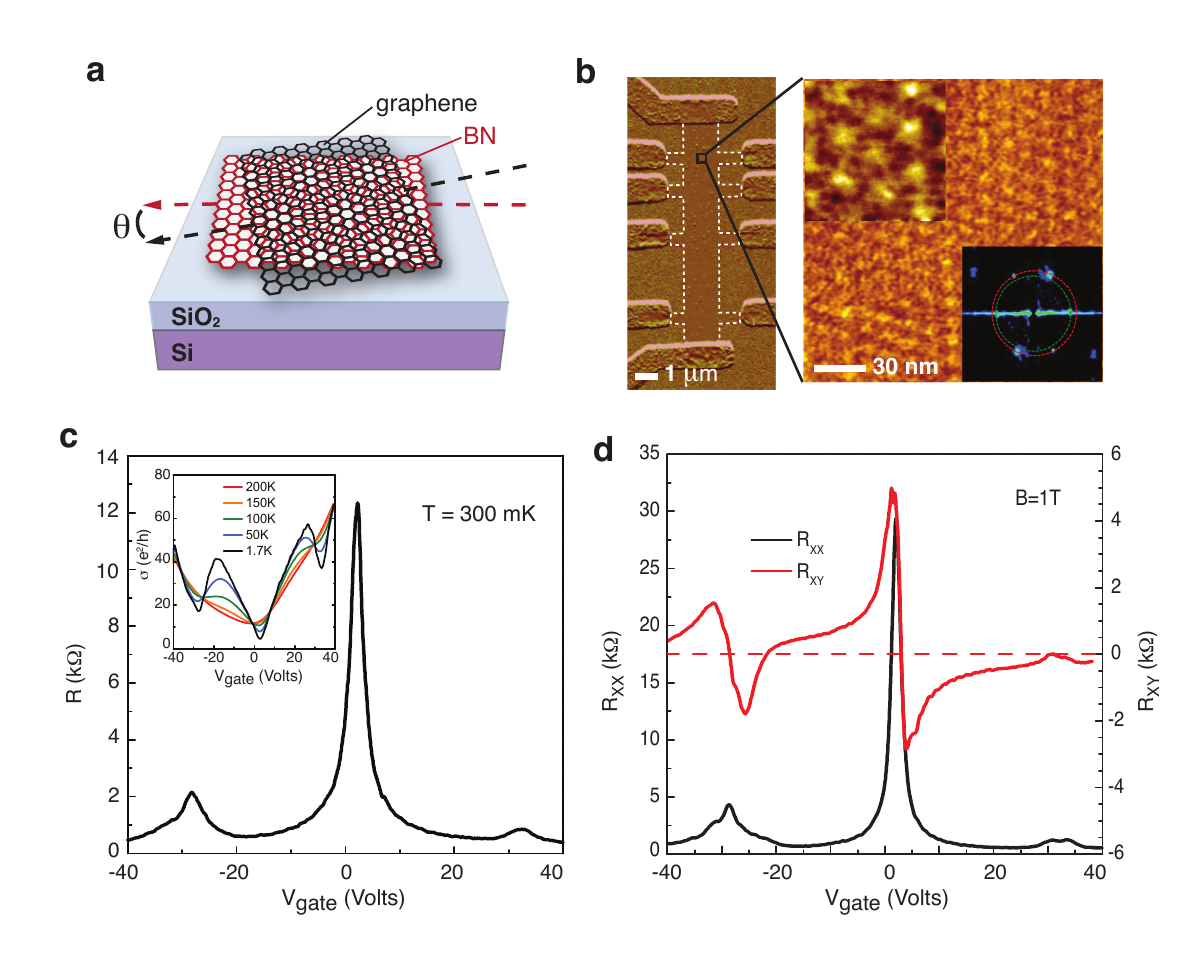}
		\caption{\textbf{Moir\'{e} superlattice} (a) Cartoon schematic of graphene on hBN showing the emergence of a moir\'{e} pattern.  The moir\'{e} wavelength  varies with the mismatch angle, $\theta$. (b) Left shows an AFM image of a multi-terminal Hall bar. Right shows a high resolution image in a magnified region.  The moir\'{e} pattern is evident as a triangular lattice (upper inset shows a further magnified region).  FFT of the scan area (lower inset) confirms a triangular lattice symmetry with period $15.5\pm0.9$~nm. (c) Resistance measured versus gate voltage at zero magnetic field.  Inset shows the corresponding conductivity versus temperature, indicating that the satellite features disappear above $\sim$100~K.  (d) Longitudinal resistance (left axis) and Hall resistance (right axis) versus gate voltage at $B=1$~T.  The Hall resistance inverts sign and passes through zero at the same gate voltage as the satellite peaks.}
		\label{fig:ZeroField}
	\end{center}
\end{figure}

\begin{figure*}[t]
	\begin{center}
	\includegraphics[width=0.9\linewidth,angle=0,clip]{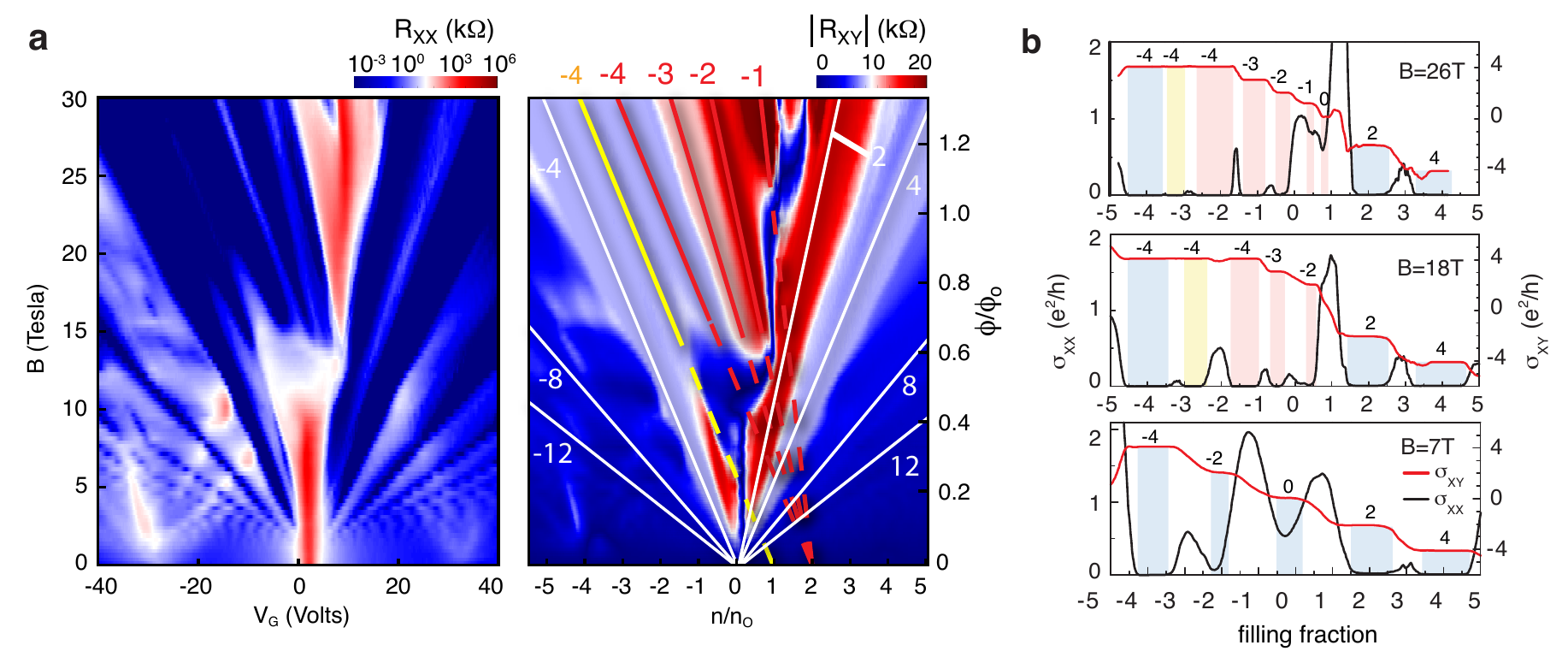}
		\caption{\textbf{Emergence of anomalous quantum Hall states} (a) Landau fan diagrams showing longitudinal resistance, $R_{xx}$, (left) and Hall resistance, $R_{xy}$ (right). $R_{xx}$ is plotted against magnetic field on the vertical and gate bias on the horizontal. In $R_{xy}$ the axes are scaled by the size of the moir\'{e} unit cell to give $\phi/\phi_{o}$ on the vertical and $n/n_{o}$ on the horizontal. QHE states corresponding to the conventional BLG spectrum are indicated by white lines. Solid yellow and red lines track QHE outside this conventional spectrum, with dashed lines indicating the projected $n/no_{o}$ intercept.  The slope of each line is labeled across the top axis.  (b) Longitudinal and transverse Hall conductivities corresponding to line cuts at constant magnetic field (constant $\phi/\phi_{o}$) from the Landau fan diagram in (a). At $B=7$~T the QHE ladder is consistent with previous reports for bilayer graphene.  At $B=18$~T and 26~T additional QHE states emerge showing Hall conductivity plateaus quantized to integer multiples of $e^{2}/h$, but appearing at non-integer LL filling fractions. Yellow and Red colored bars indicate correspondence to the same anomalous features marked by solid lines in (a). Blue bars label the conventional QHE features.  Numbers label the quantization integer for each plateau.}
		\label{fig:Device1}
	\end{center}
\end{figure*}

In this study, Bernal-stacked bilayer graphene (BLG) Hall bars are fabricated on hBN substrates (Fig.~\ref{fig:ZeroField}a,b) using mechanical exfoliation followed by co-lamination (see methods).  Fig. 1b shows a non-contact atomic force microscopy (AFM) image acquired from an example device.  In the magnified region, a moir\'{e} pattern is visible with triangular symmetry.  Fast Fourier transform (FFT) analysis of the image, shown in the inset, indicates that the moir\'{e} wavelength is $15.5\pm0.9$~nm.  This is comparable to the maximal  wavelength of $\sim14$~nm expected for graphene on hBN\cite{Decker:2011, Xue:2011, Yankowitz:2012} (set by the 1.8\% lattice mismatch between the two crystals), suggesting that in this device the BLG lattice is oriented with near zero angle mismatch to the underlying hBN lattice. 

Fig.~\ref{fig:ZeroField}c shows transport data measured from the same device.  In addition to the usual resistance peak at the charge neutrality point (CNP), occurring at gate voltage $V_{g}\sim 2$~V, two additional satellite resistance peaks appear, symmetrically located at $V_{satl}\sim\pm30$~V away from the CNP.  These satellite features are consistent with a depression in the density of states (DOS) at the superlattice Brillouin zone band edge, analogous to previous spectroscopic measurements of single layer graphene coupled to a moir\'{e} pattern\cite{Yankowitz:2012,Li:2009}. Assuming non-overlapping bands, $|V_{satl}|$, gives an estimate of the moir\'{e} wavelength to be $\sim14.6$~nm (see supplemental information), in good agreement with the AFM measurements. The nature of these satellite peaks can be further probed in the semi-classical, low $B$-field transport regime. In Fig.~\ref{fig:ZeroField}d,  longitudinal resistance, $R_{xx}$, and transverse Hall resistance, $R_{xy}$,  are plotted versus gate voltage at $B=1$~T. Near the central CNP, the Hall resistance changes sign as the Fermi energy passes from the electron to the hole band.  The same trend also appears near $V_{satl}$, consistent with the Fermi energy passing through a second band edge.  This provides further confirmation that the moir\'{e} pattern, acting as a periodic potential superlattice, gives rise to a mini-Brillouin zone band \cite{Wallbank:2012}. The satellite peaks vanish at temperatures above ~ 100 K (inset of Fig. 1c), indicating that the coupling between the BLG and hBN atomic lattices is of order $\sim$10~meV.
 
In the remainder of this letter, we focus on magnetotransport measured at high field. Fig.~\ref{fig:Device1}a shows the evolution of  $R_{xx}$ and  $R_{xy}$ for magnetic fields up to 31~T. In the left panel $R_{xx}$ is plotted against the experimentally tunable gate voltage and magnetic field (a so-called Landau fan diagram). In the right panel, the magnitude of the corresponding $R_{xy}$ is plotted against the dimensionless parameters appearing in the Diophantine equation, $n/n_{o}$ and $\phi/\phi_{o}$.  This ``Wannier diagram'' is simply the Landau fan diagram with both axes relabeled by dimensionless units defined by normalizing to the moir\'{e} unit cell area.

In a conventional quantum Hall system, the Landau fan diagram exhibits straight lines, tracking minima in $R_{xx}$ and plateaus in $R_{xy}$.  Plotted against $n/n_{o}$ and $\phi/\phi_{o}$, the slope of each line is precisely the LL filling fraction, $\nu$, and all lines converge to the origin. White lines in Fig.~\ref{fig:Device1}a identify QHE states matching this description, tracking LL filling fractions $\nu=4,8$ and $12$.  This is consistent with the usual QHE hierarchy associated with a conventional degenerate BLG spectrum.

\begin{figure*}[t]
	\begin{center}
	\includegraphics[width=0.9\linewidth,angle=0,clip]{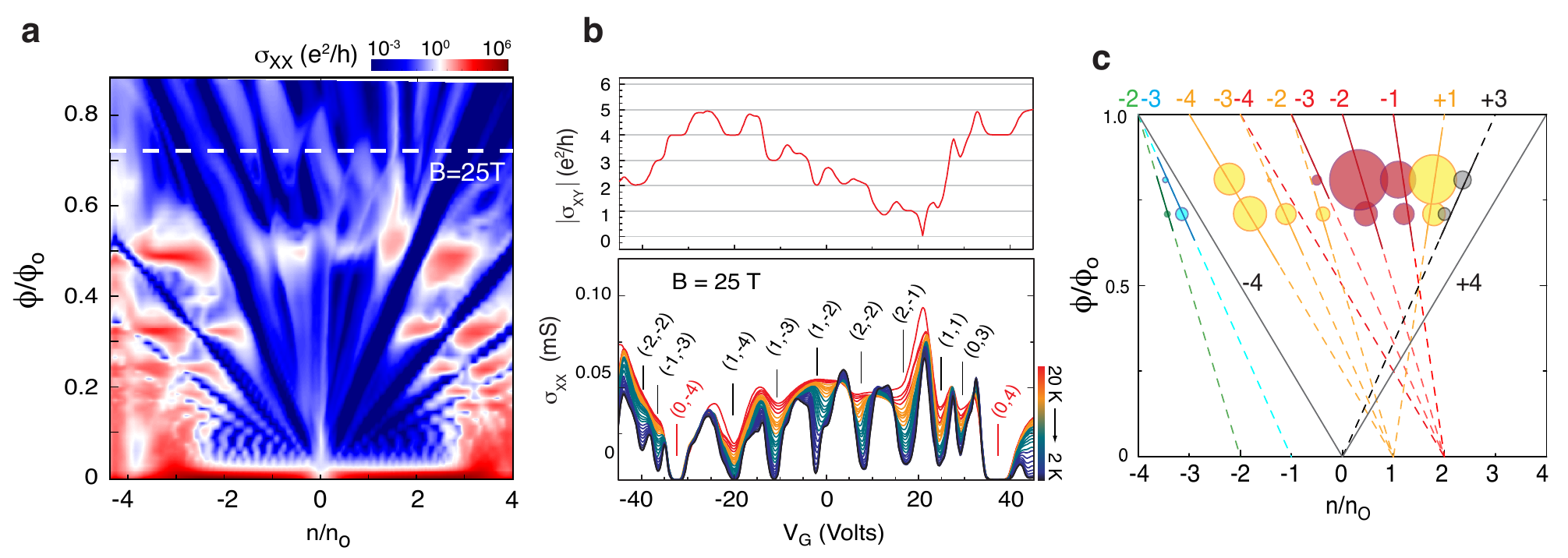}
		\caption{\textbf{Fractal gaps} (a) Landau fan diagrams similar to Fig. 2 but measured from a separate device.  Here the zero-field satellite peak position indicates  a moir\'{e} period of 11.6~nm or approximately 1.5$\times$ smaller superlattice unit cell area. Significantly more structure is observed compared with the previous device (b).  Lower panel shows the evolution of $\sigma_{xx}$ with temperature varying between 2~K and 20~K, acquired at constant $B=25$~T corresponding to the line cut shown in (a). Upper panel shows the corresponding $\sigma_{xy}$ at $T=2$~K. The bracketed numbers label the $(s,t)$ values of the corresponding fractal gaps according to the Diophantine equation. (c) Bubble plot of energy gaps determined from the temperature dependence acquired at two magnetic fields ($B=25$~T and 28.5~T).  The gaps are plotted as circles with radius scaled relative to the largest gap value measured. Dashed lines trace out select fractal gap positions allowed by the Diophantine equation.  Solid lines trace regions where the corresponding fractal gaps appear as minimia in $\sigma_{xx}$ together with quantized plateaus in $\sigma_{xy}$.}
		\label{fig:Device2}
	\end{center}
\end{figure*}

 At large magnetic fields, several additional QHE states, exhibiting minima in $R_{xx}$ together with plateaus in $R_{xy}$, develop outside the usual BLG sequence that also follow straight lines in the Landau fan diagram but converge to non-zero values of $n/n_{o}$. Yellow and red lines in Fig.~\ref{fig:Device1}a trace examples of these anomalous QHE states appearing within the lowest LL.  Unlike the conventional QHE states, each of the anomalous QHE states is characterized by both an integer valued intercept, $s$ (yellow and red lines converge to $n/n_{o}=1$ and 2, respectively) and an integer valued slope, $t$ (labeled along the top axis in the figure).  In Fig.~\ref{fig:Device1}b, longitudinal and Hall conductivities acquired at constant magnetic field (corresponding to horizontal line cuts through the fan diagram in Fig.~\ref{fig:Device1}a) are plotted against LL filling fraction, $\nu$. At large magnetic fields the anomalous QHE states are remarkably well developed, exhibiting wide plateaus in $\sigma_{xy}$ concomitant with zero valued $\sigma_{xx}$. Moreover, these states appear in general at non-integer filling fractions. Comparison between Fig.~\ref{fig:Device1}a and \ref{fig:Device1}b further reveals that Hall conductivity plateaus are quantized to integer values $t e^{2}/h$, where the quantization integer $t$ equals the slope in the Wannier diagram. Similar internal structure is observed within higher order 
Landau levels (see Fig.~\ref{fig:Device2} and also supplemental info). The anomalous QHE states observed here are consistent with fully developed spectral gaps resulting from a Hofstadter-type energy spectrum. Moreover, our ability to fully map the density-field space  provides a remarkable confirmation of the Diophantine equation, where we observe direct evidence that QHE features associated with the  Hofstadter spectral gaps are characterized by the two quantum numbers, $s$, and $t$, corresponding to the $n/n_{o}$ intercept and slope, respectively, in the Wannier diagram.

\begin{figure*}[t]
	\begin{center}
	\includegraphics[width=0.75\linewidth,angle=0,clip]{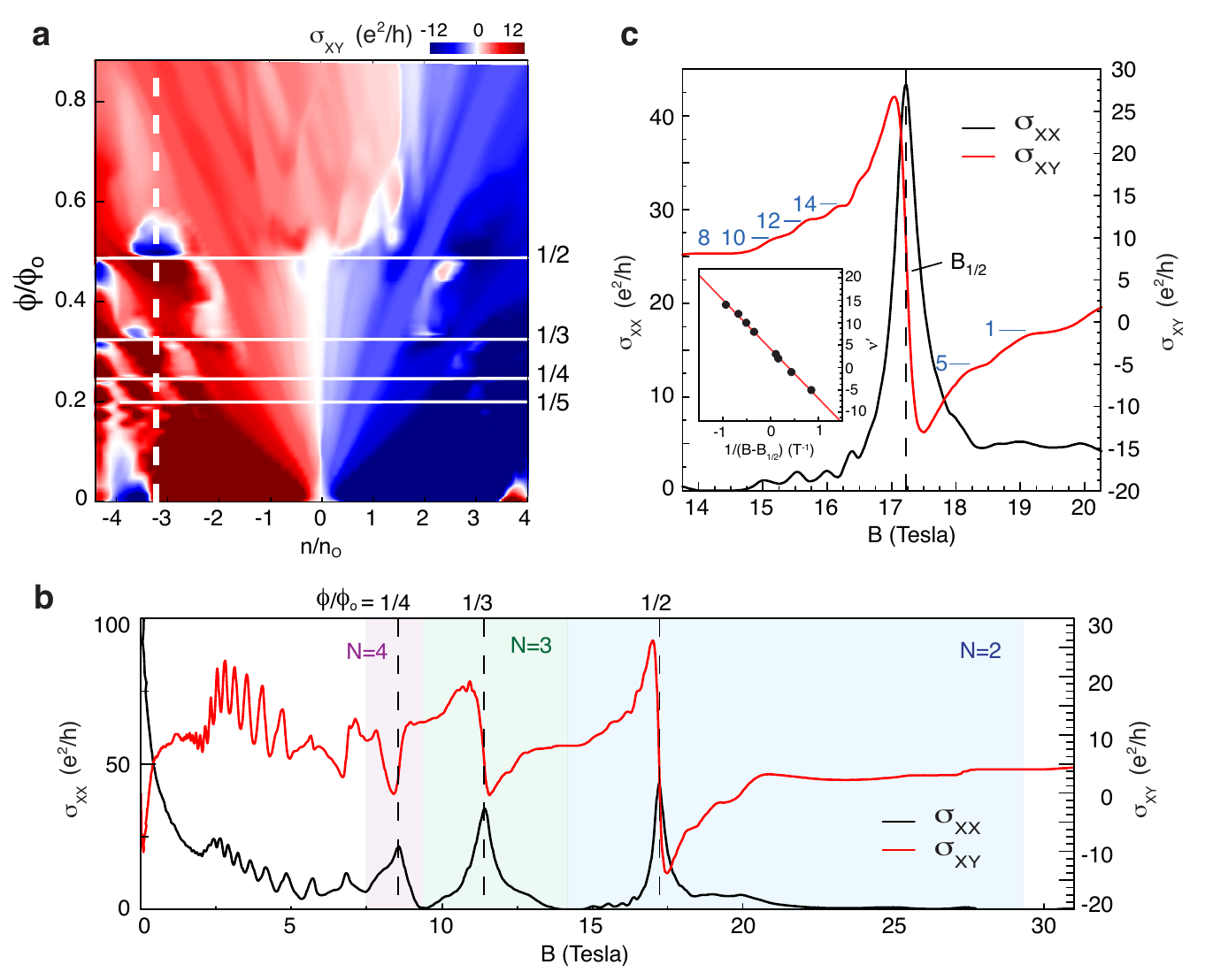}
		\caption{\textbf{Recursive structure} (a) $R_{xy}$ Wannier diagram for the same device shown in Fig. 3.  White solid lines label $\phi/\phi_{o}$ values corresponding to ``pure fractions'', $\phi/\phi_{o}=1/m$.  (b) Longitudinal conductivity, $\sigma_{xx}$ (left axis) and Hall conductivity, $\sigma_{xy}$ (right axis) versus magnetic field, acquired at a constant gate voltage ($V_{g}=-39$~V, corresponding to the white dashed line in (a).  Blue, green and purple bands mark the boundaries of the conventional $N=2,3$ and 4 Landau levels, respectively, where $N$ is the index of the orbital wavefunction. Dashed lines mark the $B$ field corresponding to the pure fractions labeled in (a). (c) Expanded plot showing details of both $\sigma_{xx}$ and $\sigma_{xy}$ in the vicinity of $\phi/\phi_{o}=1/2$ from (b). Plateaus in $\sigma_{xy}$ concomitant with minima in $\sigma_{xx}$ resemble a a mini QHE trace centered around $\phi/\phi_{o}=1/2$, consistent with Hofstadter's prediction of recursion in the butterfly spectrum. Inset shows the position of the mini QHE plateaus plotted versus effective magnetic field following a linear trend (see text) }
		\label{fig:Fractal}
	\end{center}
\end{figure*}

Fig.~\ref{fig:Device2} shows similar data to Fig.~\ref{fig:Device1}, but acquired from a separate device in which the moir\'{e} wavelength is only 11.6~nm. Again, QHE states appear outside of the conventional Bernal-stacked BLG sequence following straight lines whose origin and slope are both integer valued, and with the slope exactly matching the Hall quantization, in precise agreement with the Diophantine equation. Similar to the previous device,  the $\nu=0$ insulating state undergoes a dramatic change near $\phi/\phi_{o}=1/2$ when anomalous QHE states associated with the fractal gaps begin to fully develop. 

In Fig.~\ref{fig:Device2}b the lower panel shows $\sigma_{xx}$ acquired at $B=25$~T,  corresponding to a horizontal line cut through Fig.~\ref{fig:Device2}a, with varying sample temperature. The magnitude of the fractal gaps were estimated from the temperature dependence of the $\sigma_{xx}$ minima in the thermally activated regime (see methods).  This analysis was performed at two magnetic field values, $B=25$~T and 28.5~T. Fig.~\ref{fig:Device2}c summarizes our findings.  Each fractal gap is marked by a circle centered at the corresponding $n/n_{o}$, $\phi/\phi_{o}$ coordinate, and with radius scaled relative to the largest gap value (all gap values are given in the supplemental information). With increasing magnetic field spectral energies develop in a complicated way where some gaps grow with field (\textit{e.g.} $(s,t)=(1,1)$ and $(2,-2)$), while others diminish (\textit{e.g.} $(1,-4)$). At fixed magnetic field it appears generally true that for constant $s$ values, fractal gap states with increasing $t$ exhibit increasing gap size. For example, at $B=25$~T, $\Delta_{(1,-4)}\sim48$~K, whereas $\Delta_{(1,-3)}\sim30$~K.  This contradicts Hofstadter's prediction that fractal gaps corresponding to lower quantum numbers exhibit larger gap values.  We note that such a trend was subsequently found to be specific to square lattice symmetries \cite{Wannier:1978,Claro:1979,MacDonald:1983}.  Furthermore, a non-trivial case also arises when two fractal gap states overlap\cite{Wannier:1978}, such as in our data between the (1,-2) and (2,-3) states as $\phi/\phi_{o}\rightarrow 1$.  Further theoretical analysis specific to moir\'{e}-patterned BLG is necessary to fully understand the trends highlighted here.  

Fig.~\ref{fig:Fractal}a shows a normalized Landau fan diagram of $R_{xy}$ values corresponding to the same $R_{xx}$ data in Fig.~\ref{fig:Device2}a.  Dashed horizontal lines in the figure label special values $\phi/\phi_{o}=1/m$, where $m$ is integer valued.  Referred to by Hofstadter as the ``pure cases''\cite{Hofstadter:1976}, these lines of high symmetry provide the framework for the recursive structure of the butterfly spectrum, marking the boundaries of the repeating sub-cells that appear within the main cell\cite{MacDonald:1983}.  It appears that at particular points in the fan diagram, $R_{xy}$ tends towards zero, inverting sign as the magnetic field is swept through these pure fractions. This is further revealed by the single line-trace in in Fig.~\ref{fig:Fractal}b. At low magnetic fields, we observe the semi-classical Weiss oscillations previously reported for square superlattices\cite{Weiss:1990,Gerhardts:1991,Schlosser:1999,Albrecht:1998}. In the quantum Hall regime, the longitudinal conductivity shows a local peak as the magnetic field passes through the pure fractions, with the corresponding Hall conductivity exhibiting a sharp transition. Near the field corresponding to  $\phi/\phi_{o}=1/2$, labeled in Fig.~\ref{fig:Fractal}c as $B_{1/2}$, plateaus appear in $R_{xy}$ together with minima in $R_{xx}$ resembling a mini QHE series centered around $B_{1/2}$. If we redefine the local effective magnetic field as $B'=B-B_{1/2}$, then according to the usual QHE formalism we expect the relation $\nu'=(1/B')n'h/e$, where $\nu'$ is an effective filling fraction given by the Hall quantization, $B'$ labels the effective magnetic field position of the $R_{xx}$ minima, and $n'$ represents an effective carrier density.  Inset in Fig.~\ref{fig:Fractal}c shows a plot of $\nu'$ versus $1/B'$ and indeed, the data follows a linear trend.  In spite of the large magnetic field ($B_{1/2}\sim17.3$~T), this remarkable observation indicates that locally the electrons behave as if the magnetic field is reduced to zero. We regard this as compelling evidence of the long-predicted recursive nature of the Hofstadter spectrum where repeated mini fan diagrams emerge within the main one. 
Interestingly, the linear trend shown inset in Fig.~\ref{fig:Fractal}c does not pass through the origin, but is vertically offset by $4.1\pm0.1$.  The origin of this offset is unclear but may be related to disorder effects since in this regime the spectrum is not fully gapped\cite{Koshino:2006}.

Finally, we discuss the apparent broken symmetries observed in our data. We first note that the most prominent fractal gap states within the lowest Landau level of both devices correspond to positive $s$, but negative $t$ quantum numbers. Likewise, the overall fractal gap structure appears stronger on the hole side (negative density values) than the electron side (positive density values).  This electron/hole asymmetry results from coupling between BLG and hBN, which, in a single-particle tight binding calculation, tends to break the bipartite nature of the BLG lattice. We further note that our data exhibits complete symmetry breaking of both spin and pseudospin degeneracy, with fractal gaps appearing for both even and odd values of $s$ and $t$, whereas only multiples of 4 may be expected in a fully degenerate case. The pseudospin degeneracy, which is related to the spacial inversion symmetry, is lifted due to asymmetric coupling to the substrate, where the bottom graphene layer interacts more strongly with the BN substrate than the top graphene layer. In the high magnetic field regime, Zeeman coupling may break spin symmetry to yield odd integer QHE states. A  detailed consideration of these effects can be found in the supplemental information. However, we note that at these fields the Coulomb energy is more than an order of magnitude larger than Zeeman\cite{Dean:2011}, suggesting that interaction-induced spontaneous symmetry breaking of the internal quantum degrees of freedom may be necessary to understand our experimental observations.

\section*{Methods}
Devices were fabricated using a co-lamination mechanical transfer technique similar to that described previously\cite{Dean:2010,Dean:2011}. Electrical leads consisting of a Cr/Pd/Au metal stack were deposited using standard electron-beam lithography after which the sample was etched into a  Hall bar by exposure to oxygen plasma. Graphene/hBN stacks were fabricated on doped Si substrates with a $\sim300$~nm oxide layer. More than 20 devices were made in similar way, where 6 devices show similar behavior to that reported here. We focus only on 2 high quality devices in the text, listing other examples in the supplementary materials.  In electrical measurements the charge carrier density was varied using the doped silicon as a field effect gate. Four-terminal transport measurements were performed using a lock-in amplifier at 17~Hz with a $10-100$~nA source current.  Samples were measured in a 31~T resistive magnet and $^{3}$He cryostat (sample in vapour). Longitudinal and Hall conductivities were calculated from the corresponding measured resistance according to $\sigma_{xx}=\rho_{xx}/(\rho_{xx}^{2}+R_{xy}^{2})$ and  $\sigma_{xy}=R_{xy}/(\rho_{xx}^{2}+R_{xy}^{2})$, respectively.

AFM images were acquired after device fabrication was complete, using an Omicron low temperature AFM system operated at room temperature. Imaging was performed using $V_{bias} = 0.2$~V and $\delta f=20$~Hz. Images were filtered to remove noise.

Gap energies were estimated from the temperature dependence of longitudinal conductivity minima in the thermally activated regime, 
$\sigma_{xx}\propto e^{\Delta/2k_{B}T}$ where $\Delta$ is the energy gap, $k_{B}$ is Boltzmann's constant and $T$ is the electron temperature. Each gap value was determined from the corresponding Arrhenius plot by fitting to the linear regime.

\bigskip
\section*{Acknowledgments}
We thank A. MacDonald for helpful discussions.  A portion of this work was performed at the National High Magnetic Field Laboratory, which is supported by National Science Foundation Cooperative Agreement No. DMR-0654118, the State of Florida and the U.S. Department of Energy.  This work is supported by AFOSR MURI, FCRP through C2S2 and FENA. PK and FG acknowledge sole support from DOE (DE-FG02-05ER46215). JK and MI were supported by by the National Science Foundation under Grant No. 0955625.

%\bibliographystyle{naturemag}
%\bibliography{Refs}

\clearpage

\setcounter{figure}{0}
\makeatletter \renewcommand{\thefigure}{S\@arabic\c@figure} \renewcommand{\thetable}{S\@arabic\c@table} \makeatother %Label figures "Sxx"

\section*{Supplementary Information}

%%%%%%%%%%%%%%%%%%%%%%

\section{Wannier diagram}

\begin{figure}[h]
	\begin{center}
	\includegraphics[width=1\linewidth,angle=0,clip=]{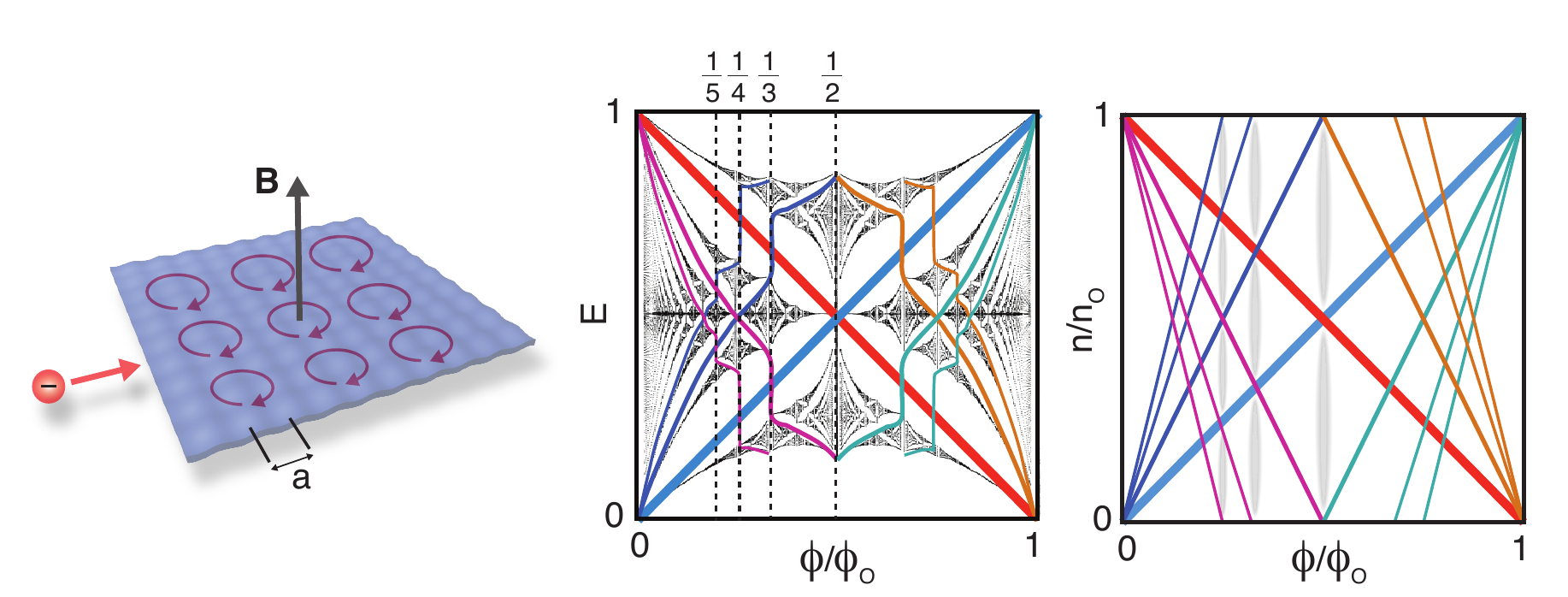}
		 \caption{Left is a cartoon image of an electron subjected to both a magnetic field, and a square periodic lattice.  The usual Hofstadter Butterfly spectrum, calculated for a square superlattice, is shown in the middle. Right shows the spectrum replotted as density versus magnetic field.  Coloured lines in the Hofstadter energy-field spectrum (middle) follow lines of constant chemical potential.  In the Wannier density-field spectrum (right), these same lines follow linear trajectories, according to the Diophantine equation.}
		\label{fig:Hofstadter}
	\end{center}
\end{figure}

Fig. \ref{fig:Hofstadter}b and \ref{fig:Hofstadter}c illustrates the relationship between the energy-field diagram and density-field diagram.  As first demonstrated by Wannier (see main text), all spectral gaps follow linear trajectories in the density-field space according to a dimensionless Diophantine equation (Eqn.~1 in the main text). The Wannier diagram is experimentally accessible by performing transport measurement while varying the carrier density and magnetic field.  In the quantum Hall regime, spectral gaps in the spectrum appear as minima in longitudinal resistance, $R_{xx}$, and quantized plateaus in the transverse Hall resistance, $R_{xy}$.

%%%%%%%%%%%%%%%%%%%%

\section{Replotting Landau fan diagram in dimensionless units}

\begin{figure}[h]
	\begin{center}
	\includegraphics[width=0.5\linewidth,angle=0,clip=]{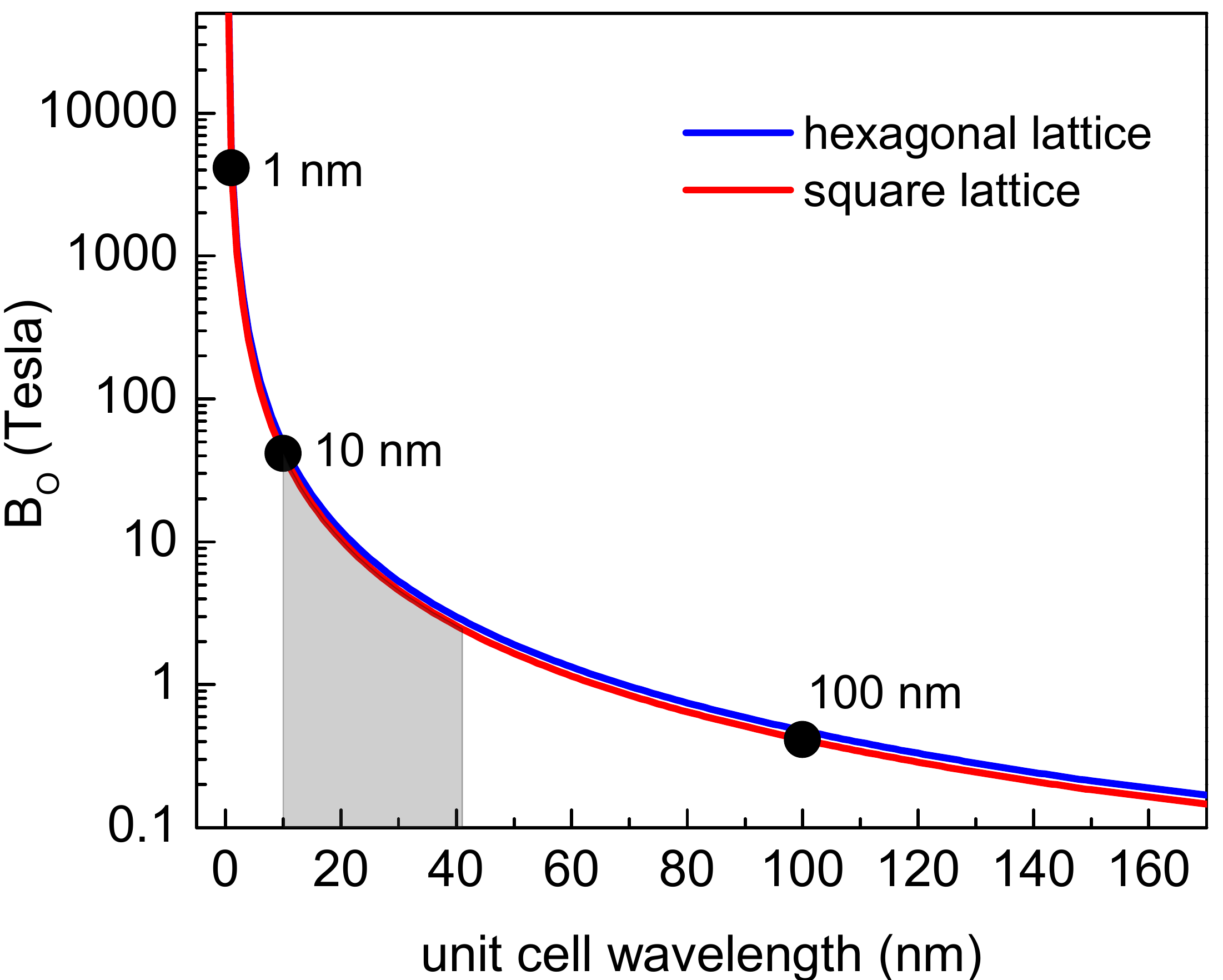}
		 \caption{Critical field versus unit cell length, defined where the ratio $\phi/\phi_{o}\rightarrow1$. Gray shaded area highlights the ideal unit cell length for magnetic fields achievable in the laboratory.}
		\label{fig:BvsA}
	\end{center}
\end{figure}

The relevant quantities in the Landau fan diagram for a 2D electron system with a superimposed periodic potential are the size of the superlattice unit cell, and the number of flux penetrating this unit cell. For a square superlattice the magnetic flux penetrating the unit cell is

\begin{equation}
\label{eqn:PhiPhioSquare}
\frac{\phi}{\phi_{o}}=\frac{Ba^2}{\phi_{o}}
\end{equation}

\noindent
where, $B$ is the applied magnetic field, $a^2$ is the unit cell area, and $\phi_{o}=h/e$ is the magnetic flux quantum.  For a hexagonal superlattice, the unit cell area is $\frac{\sqrt{3}}{2}a^2$ where $a$ is the unit cell basis vector, giving

\begin{equation}
\label{eqn:PhiPhioHex}
\frac{\phi}{\phi_{o}}=\frac{eB\sqrt{3}a^2}{2h}
\end{equation}

\noindent
We also normalize the electron density by the inverse area of the unit cell, i.e., $n\rightarrow n/n_{o}$, where

\begin {equation}
\label{eqn:no}
n_{o}=\frac{1}{unit\,cell\,area}=\frac{1}{\frac{\sqrt{3}}{2}a^2}
\end{equation}

\noindent
We can define the critical field as the field were $\phi/\phi_{o}\rightarrow 1$, \textit{i.e.} when the magnetic length approaches the scale were there is approximately 1 flux quantum per unit cell area.   From \ref{eqn:PhiPhioSquare} and \ref{eqn:PhiPhioHex}, this gives the corresponding critical magnetic field

\begin{equation} 
\label{eqn:Bcrit}
%B_{o} &= \frac{h}{ea^{2}} &\text{square lattice,}\\
B_{o}= \frac{2h}{\sqrt{3}ea^{2}}
\end{equation}

\noindent
The same analysis for a square lattice gives $B_o=h/(ea^2)$. Fig.~\ref{fig:BvsA} shows the critical field versus unit cell wavelength.  The ideal unit cell length is highlighted in the grey shaded region, spanning the magnetic field range from a few Tesla up to approximately 40 Tesla.

\section{Moir\'{e} superlattices in bilayer graphene on hBN}

\begin{figure}[h]
	\begin{center}
	\includegraphics[width=1\linewidth,angle=0,clip=]{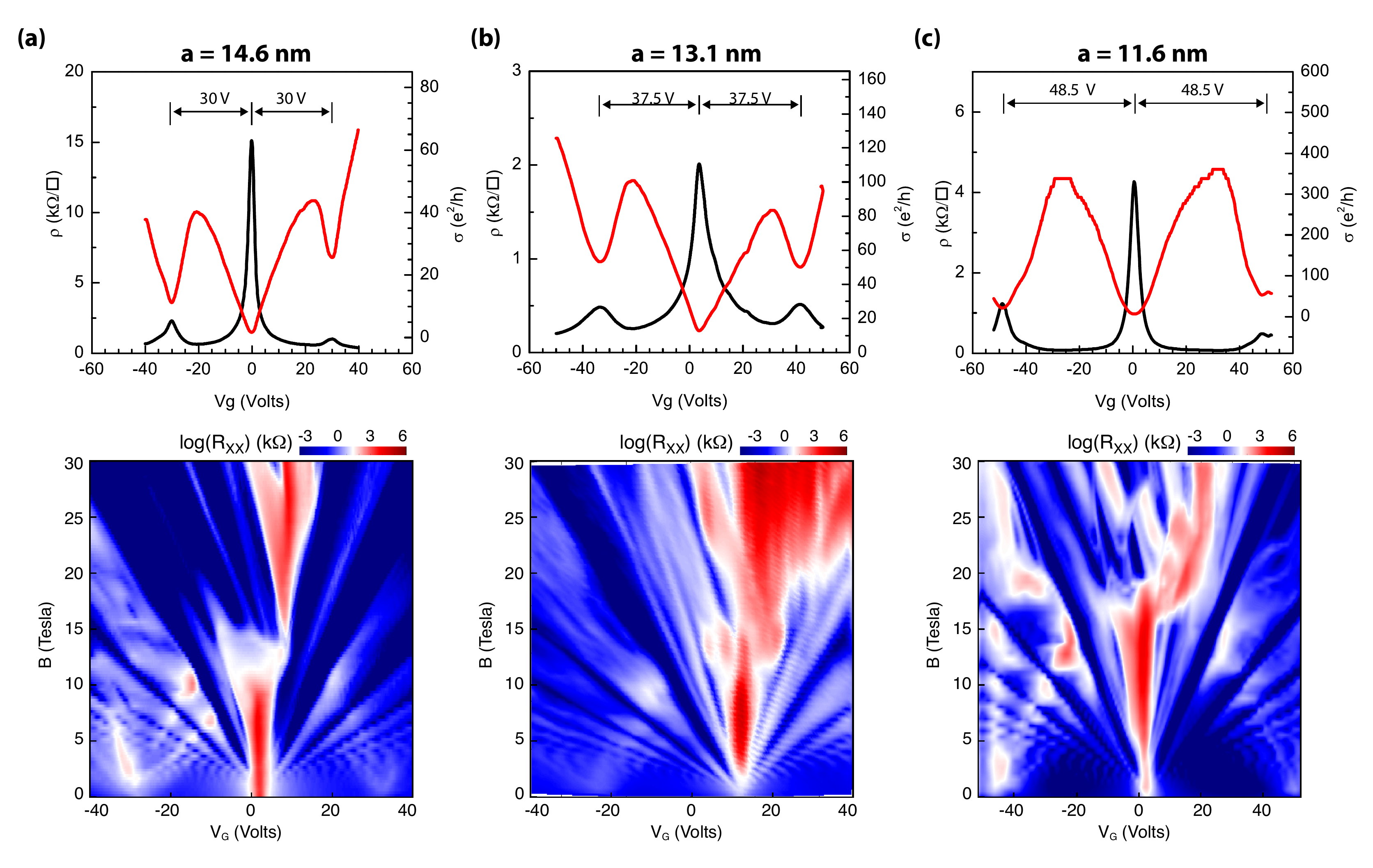}
		 \caption{(a)-(c) shows transport measured at zero magnetic field (top panel) and corresponding $R_{xx}$ plotted in a Landau fan diagram (bottom panel), for three different devices, which we label here S1,S2 and s3. In each panel, $a$ indicates the corresponding moir\'{e} wavelength, determined from the position of the satellite resistance peaks at zero field (see text).}
		\label{fig:3peaks}
	\end{center}
\end{figure}

Fig. \ref{fig:3peaks} shows three different bilayer graphene/hBN devices that exhibit modulated transport due to the presence of a moir\'{e} pattern. Device s1 (Fig.~\ref{fig:3peaks}a) corresponds to the device in Fig. 2 of the main text, with the longest moir\'{e} wavelength observed in our devices, s3 (Fig.~\ref{fig:3peaks}c) corresponds to Fig. 3 of the main text, with the shortest moir\'{e} wavelength amongst our devices. Device s2 (Fig.~\ref{fig:3peaks}b) shows a device with an intermediate moir\'{e} wavelength.   Assuming a filled band model the moir\'{e} wavelength was determined from the satellite peak position according to 

\begin{equation}
\label{eqn:Density}
\frac{n_{sat}}{n_{o}}=g_{s}g_{v}
\end{equation}

\noindent
where $n_{sat}$ is the field-effect density at the satellite peak position, $n_{o}=1/A$ where $A=\sqrt{3}a^{2}/2$ is the unit cell area of the moir\'{e} pattern (assuming hexagonal symmetry; see Fig. 1 in the main text) with $a$ the moir\'{e} wavelength, $g_{s}$ is the electron spin degeneracy and $g_{v}$ is the valley, or pseudospin, degeneracy. $n_{sat}$ is calculated according to a standard parallel plate capacitor model to be $n_{sat}=C_{g}(V_{sat}-V_{CNP})/e$, where $C_{g}$ is the geometric capacitance, $V_{sat}$ is the gate voltage corresponding to the position of the satellite resistance peak and $V_{CNP}$ is the gate voltage corresponding the central charge neutrality peak.  Substituting back into equation \ref{eqn:Density} and solving for the moir\'{e} wavelength gives

\begin{equation}
\label{eqn:Length}
a=\sqrt{\frac{8e}{\sqrt{3}C_{g}(V_{sat}-V_{CNP})}}
\end{equation}

For the three devices in Fig \ref{fig:3peaks}, the satellite peaks appear at 30~V, ,37.5~V, and 48.5~V, giving moir\'{e} wavelengths 14.6~nm, 13.1~nm and 11.6~nm, respectively.

%\begin{figure}[t!]
%	\begin{center}
%	\includegraphics[width=1.0\linewidth,angle=0,keepaspectratio=true,clip=]{Moir\'{e}VsGate.pdf}
%		 \caption{(a) Moir\'{e} lattice constant versus position in gate voltage where satellite peak emerges (equation (\ref{eqn:a(V)})).  The range in back gate where evidence of the supperlatice potential is observable is shaded in red.  The lower bound is set by the maximum possible Moir\'{e} unit cell lattice vector (peak in (b)); the upper bound is set by the maximum gate voltage we can safely apply before we typically observe onset of breakdown in the SiO2 dielectric. Red circles indicate satellite peak positions we have observed in four different samples.  (b) Moir\'{e} lattice constant, $a$, versus rotation angle between the graphene and BN layers, $\theta$ (equation (\ref{eqn:a(theta)})). The shaded region in (b) shows the range of corresponding rotational mismatch angles over which we could observe satellite peaks in transport, spanning $\sim-1.2^{o}$ to $+1.2^{o}$ }
%		\label{fig:Moir\'{e}VsGate}
%	\end{center}
%\end{figure}

\section{Theoretical Spectrum for bilayer graphene on {h}BN}

\begin{figure}[t!]
	\begin{center}
	\includegraphics[width=1\linewidth,angle=0,keepaspectratio=true,clip=]{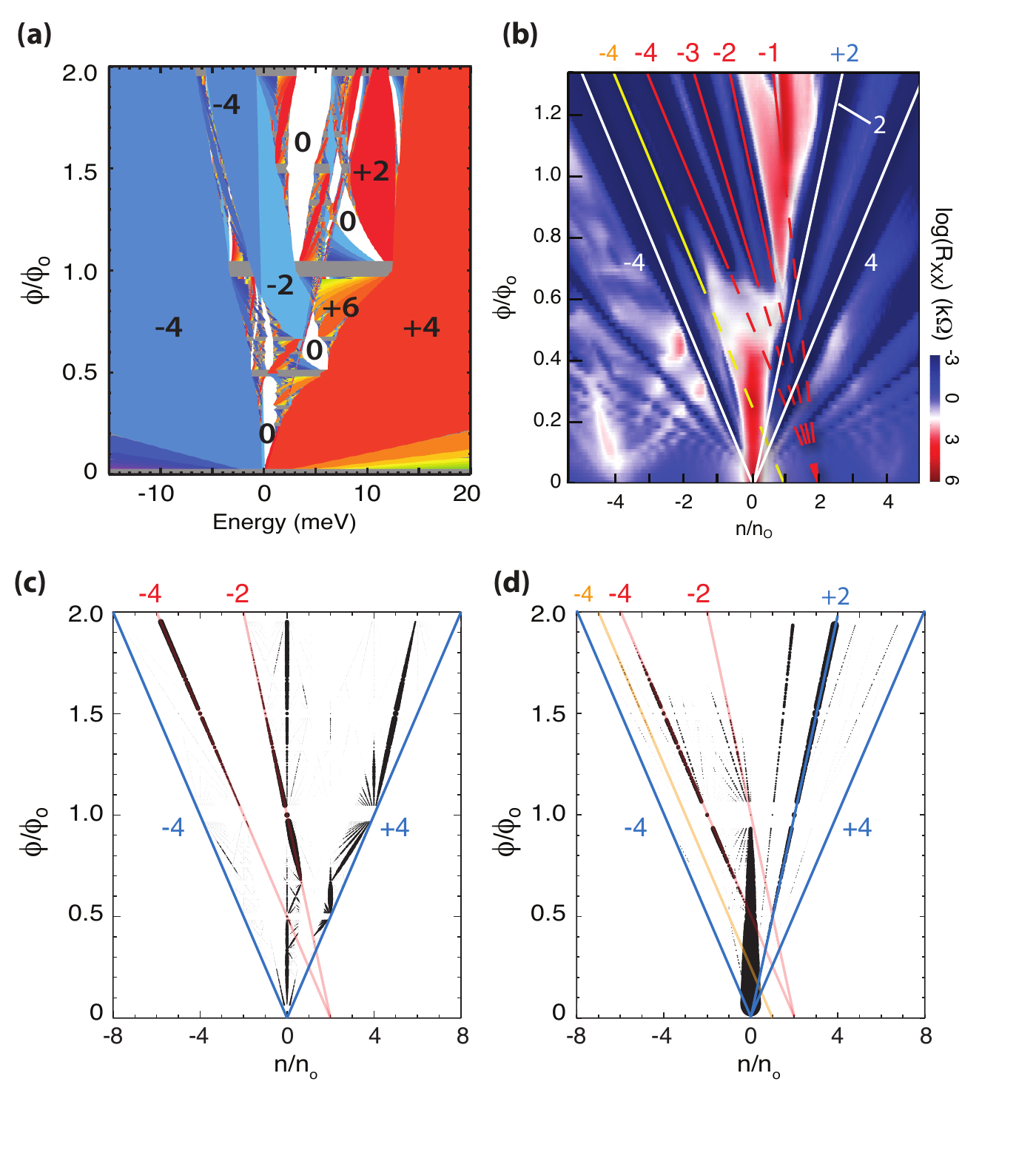}
		 \caption{(a) Energy spectrum calculated for bernal-stacked bilayer graphene on hBN with a 0$^\circ$ mismatch angle. The energy scale is chosen to highlight the lowest Landau level. The numbers label the quantized Hall conductivity in units $e^{2}/h$ corresponding to each gap. (b) Experimental magnetoresistance measured for device with similar angular mismatch (data is the same as shown in Fig. 2 in the main text). Yellow and red lines correspond to  $s=1$ and $s=2$, respectively.  Numbers along top axis label corresponding $t$ values.  (c) Wannier diagram calculated for the energy spectrum in (a), showing only gaps within the lowest Landau level.   Horizontal and vertical axes are density and magnetic flux, respectively, normalized to the area of the moir\'{e} unit cell. Fractal gaps are plotted as black circles where the radius is proportional to  the gap width. Coloured lines highlight gaps that are observed in the experimental data.  (d) Similar calculation to (b) but including Zeeman coupling and interlayer asymmetry (see text) }
		\label{fig:Theory}
	\end{center}
\end{figure}

The Diophantine equation represents a universal set of constraints describing all possible gaps within the Hofstadter spectrum. Which gaps
remain open for a particular device depends critically on the details of the system. Fig. S4a shows the calculated Hofstadter spectrum
for Bernal-stacked bilayer graphene oriented to hBN. Here graphene and hBN are modeled by tight-binding 
honeycomb lattices with the lattice periods  $a \approx 0.246$~nm and $a_{\rm hBN} \approx 0.2504$ nm,
\cite{Liu_et_al_2003a} respectively. We assume that Bernal graphene bilayer and  hBN monolayer are aligned with zero rotation angle,
and the ratio between the two lattice constants is round to  a rational number $a_{\rm hBN}/a =  56/55$ to give a finite moire superlattice period $56a \approx 13.8$~nm.  We assume the interlayer distance of bilayer grpahene to be 0.335~nm and that between hBN and graphene 0.322~nm \cite{Giovannetti_et_al_2007a}. We consider only a $p$ orbital state on each atomic site, and set on-site potential to 0, 3.34~eV and $-1.40$~eV for C, B and N atoms, respectively. \cite{Slawinska_et_al_2010a} For the hopping amplitudes between different sites,
we adopt the Slater-Koster parametrization used for twisted bilayer graphene \cite{Moon_and_Koshino_2012a}, 
irrespective of combination of atoms. To compute the low-energy spectrum in magnetic field, 
we take the low-lying Landau levels ($|E| < 1.5$~eV) of isolated bilayer graphene as the basis,  \cite{Moon_and_Koshino_2012a}
and the coupling with hBN states in the high-energy region is included as on-site potential on the bottom graphene layer (faced to hBN)
in the second-order perturbation.

In Fig. S4b, the energy spectrum is replotted as a Wannier type diagram where the energy axis is replaced by the normalized density. Fractal gaps
 inside the lowest Landau levels (between $\nu = \pm 4$) are plotted as black circles with radius proportional to the gap size. The theoretical calculation correctly predicts the experimentally observed asymmetry in the lowest Landau level (see also Fig. 3 in the main text) where fractal gaps originating from $s>0$ are much stronger than $s<0$. Related, the calculation correctly identifies
a partial lifting of the 4-fold degeneracy in graphene, evident as features associated with $s= \pm 2$, rather than only multiples of $s=4$.
 This results from broken valley degeneracy due to inversion-asymmetric coupling where the bottom graphene layer
interacts more strongly with the BN substrate than the top graphene layer. Several details in the theoretical spectrum disagree with experimental observations. For example, strong mini-gaps, predicted on the electron side, are not apparent in the experiment. Additionally, in experiment all symmetries are broken at high field, with fractal gaps corresponding to both odd and even integer values of $s$ and $t$ appearing, whereas the theoretical spectrum remains two-fold degenerate at all fields  since the spin Zeeman  splitting is neglected.

In Fig. S4c, the Wannier diagram is recalculated including spin Zeeman splitting ($g=2$) and interlayer potential
difference between the top and the bottom layer of bilayer graphene. The latter is assumed to be the sum of the constant term $30$~meV
and the gate-induced term proportional to electron density with the relative dielectric constant $\vare_r = 6$. To get a better agreement,
we also magnified coupling strength between graphene and hBN by a factor about 1.4 compared to the original (Fig. S4c).
 While additional symmetry breaking states appear, exact correspondence with our data is difficult to achieve. The model depends sensitively on physical parameters,  such as interlayer coupling strength and interlayer potential difference, that are not
well known. Moreover, this model does not account for many-body interaction, which may play a substantial role at these large magnetic
fields.  Further theoretical and experimental work is necessary to better understand the details of the observerd fractal energy spectrum.

\section{Temperature dependence}

\begin{figure}[h]
	\begin{center}
	\includegraphics[width=1\linewidth,angle=0,keepaspectratio=true,clip=]{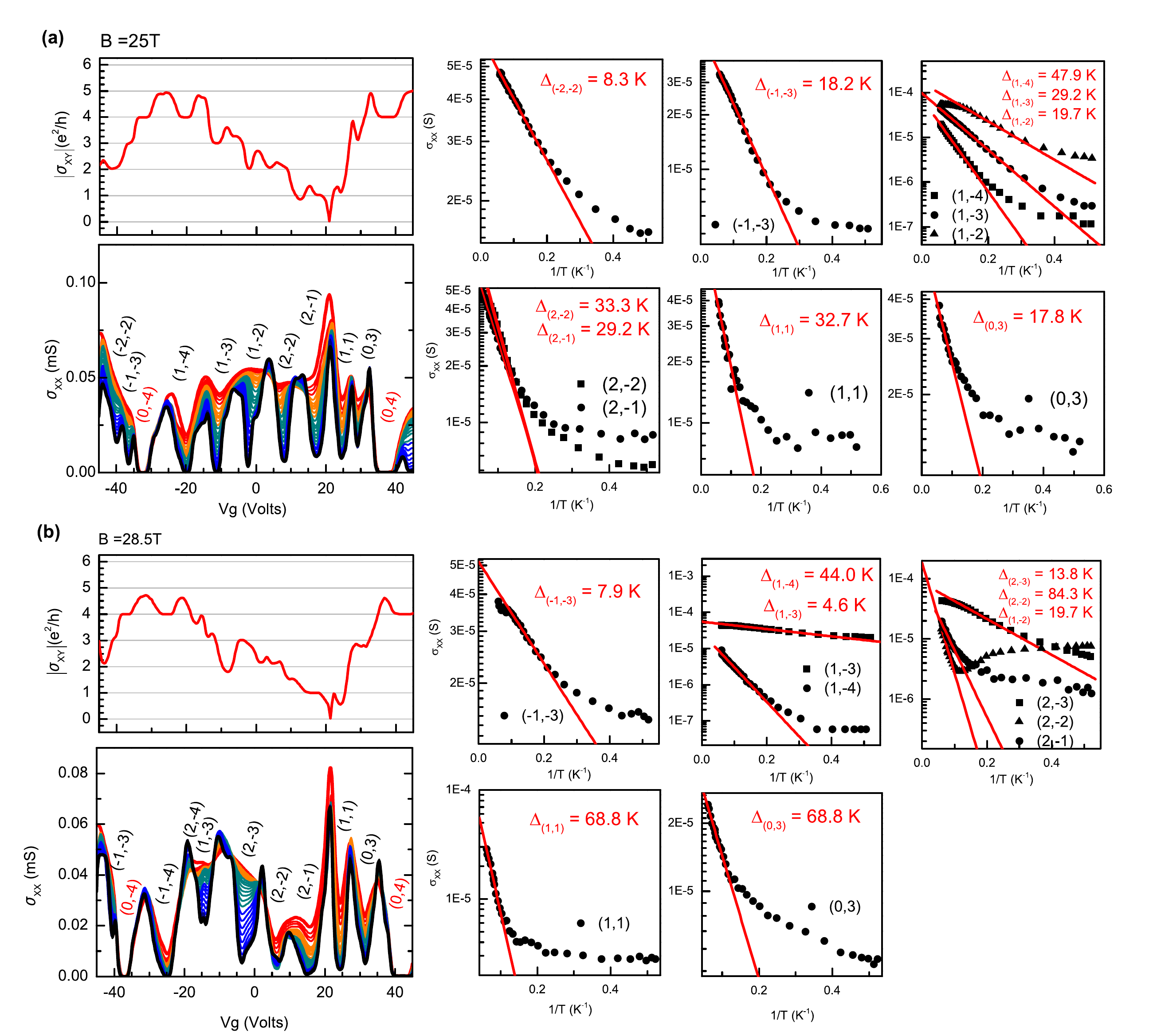}
		 \caption{Temperature dependent transport acquired at (a) $B=25$~T and (b) $B=28.5$~T.  In both (a) and (b) left panel shows the longitudinal conductivity with temperature varying between 2~K and 20~K, together with Hall conductivity at 2~K. Right panel shows Arrhenius plots for each of the fractal gaps labeled in the left panel. The corresponding gap value, determined from the slope in the linear regime, is given in each plot. }
		\label{fig:Tdep}
	\end{center}
\end{figure}

\newpage
%%%%%%%%%%%%%%

\end{document}